\documentclass[graybox]{svmult}
\usepackage{mathptmx}       % selects Times Roman as basic font
\usepackage{helvet}         % selects Helvetica as sans-serif font
\usepackage{courier}        % selects Courier as typewriter font
\usepackage{makeidx}         % allows index generation
\usepackage{graphicx}        % standard LaTeX graphics tool
\usepackage{multicol}        % used for the two-column index
\usepackage[bottom]{footmisc}% places footnotes at page bottom
\usepackage{amsmath,amssymb,amsfonts}        % AMS math packages
\usepackage{tabularx}
\makeindex             % used for the subject index
\usepackage{listings}
\usepackage{pmboxdraw}
\usepackage{newunicodechar}
\usepackage{fancyvrb}
\newunicodechar{└}{\textSFii}
\newunicodechar{├}{\textSFviii}
\newunicodechar{─}{\textSFx}

\DeclareFixedFont{\ttb}{T1}{txtt}{bx}{n}{12} % for bold
\DeclareFixedFont{\ttm}{T1}{txtt}{m}{n}{12}  % for normal

\usepackage{color}
\definecolor{deepblue}{rgb}{0,0,0.5}
\definecolor{deepred}{rgb}{0.6,0,0}
\definecolor{lightred}{rgb}{.8,.2,.2}
\definecolor{deepgreen}{rgb}{0,0.5,0}
\definecolor{cadabrapurple}{rgb}{.8,.42,.8}
\definecolor{cadabragrey}{rgb}{.5,.5,.5}

\newcommand\pythonstyle{\lstset{
language=Python,
basicstyle=\ttfamily\fontsize{8}{9.6}\selectfont,
morekeywords={self,Ex, ExNode, parent_rel_t},              % Add keywords here
keywordstyle=\ttb\color{deepblue}\fontsize{9}{10.8}\selectfont,
emph={MyClass,__init__,check_expression_for_object},          % Custom highlighting
emphstyle=\ttb\color{deepred}\fontsize{9}{10.8}\selectfont,,    % Custom highlighting style
stringstyle=\color{deepgreen}\scriptsize,,
frame=tb,                         % Any extra options here
showstringspaces=false
}}

\lstnewenvironment{python}[1][]
{
\pythonstyle
\lstset{#1}
}
{}

\newcommand\pythoninlinestyle{\lstset{
language=Python,
basicstyle=\ttfamily\fontsize{8.5}{10}\selectfont,
morekeywords={self, Ex, ExNode},              % Add keywords here
keywordstyle=\ttb\color{deepblue}\fontsize{9}{10.8}\selectfont,
emph={MyClass,__init__,check_expression_for_object},          % Custom highlighting
emphstyle=\ttb\color{deepred}\fontsize{9}{10.8}\selectfont,    % Custom highlighting style
stringstyle=\color{deepgreen},
frame=tb,                         % Any extra options here
showstringspaces=false
}}

\newcommand\pythoninline[1]{{\pythoninlinestyle\lstinline!#1!}}

\newcommand\cadabrastyle{\lstset{
basicstyle=\ttfamily\fontsize{8.5}{10}\selectfont,
morekeywords={LaTeXForm, Indices, Integer, KroneckerDelta, Derivative, Symmetric, TableauSymmetry, Depends, from, import},  
keywordstyle=\ttb\color{cadabrapurple}\fontsize{9}{10.8}\selectfont,
emph={CleverSimplify, substitute, distribute, eliminate_kronecker, unwrap, canonicalise, product_rule,check_expression_for_object, main, contract_dummy_auto, contract_epsilon_auto, rename_dummies, CleanUp, Ex, ExpressionSaver, save, read, check_time, sort, Sort, sort_product},          % Custom highlighting
emphstyle=\ttm\color{deepred}\fontsize{9}{10.8}\selectfont,    % Custom highlighting style
stringstyle=\color{deepgreen},
frame=none,                         % Any extra options here  
showstringspaces=false,
morecomment=[l][\color{cadabragrey}]{\#\#}
}}

\lstnewenvironment{cadabra}[1][]
{
\cadabrastyle
\lstset{#1}
}
{}

\newcommand\cadabrainlinestyle{\lstset{
basicstyle=\ttfamily\fontsize{8.5}{10}\selectfont,
morekeywords={LaTeXForm, Indices, Integer, KroneckerDelta, Derivative, Symmetric, TableauSymmetry, Depends},  
keywordstyle=\ttb\color{cadabrapurple}\fontsize{9}{10.8}\selectfont,
emph={CleverSimplify, substitute, distribute, eliminate_kronecker, unwrap, canonicalise, product_rule,check_expression_for_object, main, sort_product},          % Custom highlighting
emphstyle=\ttm\color{deepred}\fontsize{9}{10.8}\selectfont,    % Custom highlighting style
stringstyle=\color{deepgreen},
frame=none,                         % Any extra options here  
showstringspaces=false,
morecomment=[l][\color{cadabragrey}]{\#\#}
}}

\newcommand\cadabrainline[1]{{\cadabrainlinestyle\lstinline!#1!}}

\newcommand {\cA}{{\cal A}}
\newcommand {\cB}{{\cal B}}

\newcommand {\cD}{{\cal D}}

\newcommand {\cH}{{\cal H}}

\newcommand {\cN}{{\cal N}}

\def\a{\alpha}
\def\b{\beta}

\def\d{\delta}
\def\e{\epsilon}

\def\g{\gamma}
\def\G{\Gamma}

\def\k{\kappa}
\def\l{\lambda}

\def\r{\rho}

\def\t{\tau}

\def\ri{{\rm i}}

\newcommand{\vf}{\varphi}

\newcommand{\be}{\begin{equation}}
\newcommand{\ee}{\end{equation}}
\newcommand{\bea}{\begin{eqnarray}}
\newcommand{\eea}{\end{eqnarray}}

\newcommand{\ba}{\begin{array}}
\newcommand{\ea}{\end{array}}

\def\double #1{#1{\hbox{\kern-2pt $#1$}}}

\newcommand{\bsubeq}{\begin{subequations}}
\newcommand{\esubeq}{\end{subequations}}

\newcommand{\rd}{\mathrm d}
\newcommand{\eps}{{\epsilon}}

\begin{document}
\title*{Supergravity Component Reduction with Computer Algebra}
\author{Gregory Gold, Saurish Khandelwal, Gabriele Tartaglino-Mazzucchelli}
\institute{Gregory Gold, Saurish Khandelwal, Gabriele Tartaglino-Mazzucchelli
\at School of Mathematics and Physics, University of Queensland, St Lucia, Brisbane, Queensland 4072, Australia\\
  \email{g.gold@uq.edu.au, s.khandelwal@uq.edu.au, g.tartaglino-mazzucchelli@uq.edu.au}}
  
% \and Mehmet Ozkan
% \at Department of Physics, Istanbul Technical University, Maslak 34469 Istanbul, Turkey
% \email{ozkanmehm@itu.edu.tr}
% \and Peng-Ju Hu, Yi Pang
% \at Center for Joint Quantum Studies and Department of Physics,
% School of Science, Tianjin University, Tianjin 300350, China
% \email{pangyi1@tju.edu.cn, pengjuhu@tju.edu.cn}}
% \institute{Name of First Author \at Name, Address of Institute, \email{name@email.address}
% \and Name of Second Author \at Name, Address of Institute \email{name@email.address}}
\maketitle

\abstract{Using an interplay between superspace and component superconformal tensor calculus techniques, recently, the off-shell construction of the supersymmetric extension of the three independent curvature-squared invariants for minimal ($\cN=1$) gauged supergravity in five dimensions (5D) was completed. A key ingredient in obtaining these results is the implementation of computer algebra algorithms. In this report, we describe how to use \cadabrainline{cadabra} to systematically study component reduction from superspace with computer algebra in the case of 5D, $\cN=1$ supergravity.
%
% Each article should be preceded by an abstract (10--15 lines long) that summarizes the content. 
% The abstract will appear online at \url{www.SpringerLink.com}.\newline\indent
% Abstracts may have more than one paragraph.
}
\section{Introduction}

 The systematic study of locally supersymmetric invariants that include higher derivatives and higher curvature terms is still unsettled. However, by an effective field theory approach, quantum corrections in string theory take the form of an infinite series of (supersymmetric) higher-derivative terms. Our insufficient understanding of these terms is linked to several open questions in string theory.

When off-shell approaches to (local) supersymmetry exist, one can, in principle, construct higher-derivative invariants of arbitrarily high order in derivatives and curvatures. The reason is that when supersymmetry closes off-shell, it is possible to use a few building blocks for supersymmetric multiplets and tensor calculus to obtain manifestly supersymmetric invariants with no modifications of the transformation rules. Off-shell superspace techniques then help to have very compact expressions to start the analysis of general supergravity-matter couplings. We refer the reader to a list of pedagogical reviews on these topics \cite{Gates:1983nr,Wess:1992cp,Buchbinder:1998qv,Freedman:2012zz,Kuzenko:2022skv,Kuzenko:2022ajd,Ozkan:2024euj}.

In the last fifteen years, a new superspace approach to the construction of off-shell invariants in matter-coupled supergravity has been employed. This goes by the name of \emph{conformal superspace}, which was first introduced by Daniel Butter for 4D, $\cN=1$ supergravity \cite{Butter:2009cp} and later extended to different space-time dimensions and amount of supersymmetries --- we refer the reader to the following references \cite{Kuzenko:2022skv,Kuzenko:2022ajd} for pedagogical reviews and a complete list of references. In the last decade, this approach has allowed to obtain many new results pushing forward a systematic analysis of higher-derivative supergravity by using off-shell techniques, as exemplified in the results in the following incomplete list of papers \cite{Butter:2010jm,Butter:2013rba,Butter:2013lta,Butter:2014xxa,Kuzenko:2015jda,Butter:2016mtk,Butter:2017jqu,Novak:2017wqc,Butter:2018wss,Gold:2023dfe,PRL,Gold:2023ykx,Casarin:2024qdn}.

Conformal superspace is in spirit very similar to the superconformal tensor calculus (see \cite{Freedman:2012zz} for a review and references) with the difference that it employs superfields rather than standard component fields \cite{Butter:2009cp,Kuzenko:2022skv,Kuzenko:2022ajd}. It has the advantage of merging both components and superfield approaches in supergravity thereby providing compact and simple building blocks for locally supersymmetric invariants. Despite its power, in many applications it remains important to also have all invariants reduced to their components. For instance, to analyse the physical sector of a matter-coupled Poincar\'e supergravity engineered in conformal superspace one might first generate the appropriate locally superconformal actions in superspace. The superfields can then be reduced to fields of their corresponding component multiplets, and finally, nonphysical symmetries within the locally superconformal algebra are removed by gauge fixing. Successively, one integrates out auxiliary fields to retrieve the desired result. There are models of supergravity in conformal superspace in which the component reduction process and other relevant operations that are prerequisite to gauge fixing are either extremely repetitive or computationally so daunting that computer algebra systems are likely mandatory. For example, this is especially true when analyzing higher-derivative supergravities, which is the main topic of the recent works \cite{Gold:2023dfe,PRL,Gold:2023ykx} of which our current paper largely refers, but the same is true for several of the works \cite{Butter:2016mtk,Butter:2017jqu,Novak:2017wqc,Butter:2018wss,Gold:2023dfe,PRL,Gold:2023ykx,Casarin:2024qdn} where analogue computer algebra softwares were largely employed.

Computer algebra systems refer to software applications that perform symbolic manipulations on mathematical objects. They are widely used in scientific computing, but are especially pivotal in the complex tensorial calculations involved in general relativity and quantum gravity \cite{MacCallum:2018}. One popular tool that has many of the tensorial capabilities needed through external packages is of course \cadabrainline{Mathematica}, but as a general-purpose commercial and high-level system, it is not fundamentally designed for the tensor manipulations required in gravitational field theories. An alternative tool designed exactly for this purpose is preferred.

\cadabrainline{Cadabra} is an open source tool with core algorithms written in \pythoninline{C++} but programmable directly in \pythoninline{python} via a wrapper \cite{peeters2007introducing,Peeters:2006kp,Peeters:2018dyg}. The expression syntax in \cadabrainline{cadabra} is also in a \cadabrainline{LaTeX} readable format, and \cadabrainline{cadabra} notebooks render the output in \cadabrainline{LaTeX} automatically. It was designed specifically with tensor field problems in mind with many needed core features such as its support for tensors with free indices, index contractions, symmetries, fermions (for the most part), and more. As exemplified in our work, any missing functionality in its core can also be created in a straightforward way by adding custom \pythoninline{python} algorithms that manipulate the object structure representing defined tensorial expressions.

In this paper, we aim to give an overview of how supergravity component reduction (and other relevant operations) are performed by employing \cadabrainline{cadabra}. Our focus is on presenting key aspects of the algorithms used to obtain the results that appeared in our recent works \cite{Gold:2023dfe,PRL,Gold:2023ykx}. We also introduce a public repository \cite{GoldGithub} that can handle component reduction, derivative degauging, consistency checks, and more in $5D$, $\mathcal{N}=1$ and $4D$, $\mathcal{N}=2$ supergravity. We refer the reader to \cite{GoldGithub} for more detail and examples of our code in use. Note that while this paper gives much of the necessary building blocks, prior knowledge of \cadabrainline{cadabra} is useful\footnote{Documentation regarding the installation and usage of \cadabrainline{cadabra} can be found on its website \url{https://cadabra.science/man.html} and its Github repository \cite{CadabraGithub}.}. First, a simple component reduction example is performed to demonstrate the setup in Section \ref{sec:cadabraReduction}, and in Section \ref{sec:cadabraOptimization}, we show an approach to optimization when generating large expressions. Testing, representation conversion, and degauging are all discussed in Sections \ref{sec:cadabraTesting}, \ref{sec:cadabraRepConvert}, and \ref{sec:cadabraDegauge}, respectively. Lastly, for the reader's convenience, an overview of the repository is given in Section \ref{sec:cadabraRepo}.

\section{Component Reduction with Cadabra} \label{sec:cadabraReduction}

We proceed with an overview of the computational methodology when reducing superfields to components in $5D$, $\cN=1$ matter-coupled supergravity. For the mathematical and theoretical background behind the superspace expressions defined here and the physical relevance and application of all resulting component expressions, we refer the reader to \cite{Gold:2023ykx, Gold:2023dfe}. For lack of space, we do not review all of the $5D$ $\cN=1$ conformal superspace building blocks of \cite{Butter:2014xxa,Gold:2023ykx, Gold:2023dfe}. Here we simply use them for technical purposes focusing on their use in computer algebra algorithms.

The locally superconformal action principle relevant here involving the product of a linear multiplet with an Abelian vector multiplet is referred to as the BF action. This plays a fundamental role in the construction of general supergravity-matter couplings, see \cite{Kugo:2000hn,Kugo:2000af,Fujita:2001kv,Kugo:2002vc,Bergshoeff:2001hc,Bergshoeff:2002qk,Bergshoeff:2004kh} for the 5D case, and it is a main building block for the  invariants of interest. In components, and in our notation, the BF action takes the form 
as in \cite{Butter:2014xxa} 
\begin{align} \label{eq:BFcadabra}
    S_{\textrm{BF}} = - \int \rd^5 x & \, e \Big( \,v_{{a}} \cH^{{a}} + W F + X_{i j} G^{i j} + 2 \l^{\a k} \vf_{\a k} - \psi_{{a }}{}^{\a}_i(\Gamma^{{a}})_{\a}{}^{\b} \vf_{\b}^i  W \nonumber \\
    & - \ri \psi_{{a}}{}^{\a}_{i} (\Gamma^{{a}})_{\a}{}^{\b} \l_{\b j} G^{i j} +  \ri \psi_{{a}}{}^{\a}_{i} (\Sigma^{{a} {b}}){}_{\a}{}^{\b} \psi_{{b} \b j} W G^{i j} \Big)~.
\end{align}
We consider the case in which the linear multiplet superfield, $G^{i j}$, and its superconformal descendants, i.e., $\varphi_{\a \k}$, $F$, and $\cH^a$, are composite of other component fields the precise structure of which determines the invariant. For example, let us take the composite linear multiplet field that can be used to construct the off-shell Weyl-squared invariant in $5D$, $\mathcal{N}=1$, \cite{Hanaki:2006pj,Butter:2014xxa}. It is composite of Weyl multiplet superfields, that is, fields of the conformal supergravity multiplet
\begin{equation} \label{eq:compositeCadabra}
    G^{i j} \rightarrow H^{i j}_{\textrm{Weyl}} := - \frac{\ri}{2}   W^{{\alpha} {\beta} {\gamma} i} W_{{\alpha} {\beta} {\gamma}}\,^{j}+\frac{3\ri}{2}   W^{{\alpha} {\beta}} X_{{\alpha} {\beta}}\,^{i j} - \frac{3 \ri}{4}   X^{{\alpha} i} X^{j}_{{\alpha}}~.
\end{equation}
This composite expression and its superconformal descendants should then be directly inserted into Eq. (\ref{eq:BFcadabra}) to construct the desired component action where its descendants are defined as follows
\bsubeq \label{eq:descCadabra}
\begin{align}
\varphi_\a^i &:= \frac{1}{3} \nabla_{\a j} G^{ij} \ , \\
F &:= \frac{\ri}{12} \nabla^{\g i} \nabla_\g^j G_{ij} = - \frac{\ri}{4} \nabla^{\g k} \varphi_{\g k} \ ,\\
\cH^{a} &:= \frac{\ri}{12} (\G^{a})^{\a \b} \nabla_{\a}^i \nabla_{\b}^j G_{ij}  ~ .
\end{align}
\esubeq
Computing these descendants clearly requires the successive action of spinor covariant derivatives $\nabla_\a^i$ (which are the superspace generalizations of $Q^i_\a$-supersymmetry generators) on the composite linear multiplet field $H^{i j}_{\textrm{Weyl}}$ in Eq. (\ref{eq:compositeCadabra}). Following normal Leibniz rule logic, this can be written as many terms comprised of spinor covariant derivatives acting on the Weyl tensor superfield $W_{\a\b}$ and its descendants that describe the standard Weyl multiplet of $5D$ $\cN=1$ conformal supergravity. This is a straightforward exercise as one may simply apply the general definitions of ``tower" operations with appropriate index contractions where each field of the Weyl multiplet has distinct properties and identities and can be trivially projected to components. For instance, the independent fields in the standard Weyl multiplet are
\bsubeq
\begin{align}
W_{\a \b} &:= W_{\a \b} \vert ~, \\
\chi^i_\a &:= \frac{3 \ri}{32} X^i_\a \vert , ~~~~ ~~~ W_{\a \b}{}_{\g}^i := W_{\a \b} \vert ~, \\
    D &:= - \frac{3}{128} Y \vert ~, ~~~ \Phi_{\a \b}{}^{i j} := - \frac{3 \ri}{4} X_{\a \b}{}^{i j} \vert ~,
\end{align}
\esubeq
where the single vertical line next to a superfield denotes the usual component projection to
$\theta = 0$. Thus, we will retrieve component expressions when all spinor covariant derivatives have been applied, and they are directly inserted into the BF action for a component model. For the purpose of this section, two of the above-mentioned tower projection definitions are given below and the complete set can be found in \cite{Gold:2023ykx}. It holds
\bsubeq \label{cadabraWdervs-a}
\begin{align}
\nabla_{\g}^i W_{\a\b} 
=& W_{\a\b\g}{}^i + \eps_{\g (\a} X_{\b)}^i \ ,  
\\
\nabla^{i}_{\a}{X^{j}_{{\beta}}} 
=& 
X_{{\alpha} {\beta}}\,^{i j}
+\frac{\ri}{8} \eps^{i j} \eps_{{\alpha} {\beta}} Y
-\frac{3\ri}{2} \eps^{i j} {\nabla}_{{\alpha}}{}^{ \rho}{W_{{\beta} {\rho}}}
-2\ri \eps^{i j} W_{{\alpha}}{}^{{\rho}} W_{{\beta} {\rho}}
\nonumber\\
&
+\frac{\ri}{2} \eps^{i j} \eps_{{\alpha} {\beta}} W^{{\gamma} {\delta}} W_{{\gamma} {\delta}}
-\frac{\ri}{2}\eps^{i j} {\nabla}_{{\beta}}{}^{{\rho}}{W_{{\alpha} {\rho}}}~,
\end{align}
\esubeq
where $\nabla_{\a}{}^{\b} = (\Gamma_a)_{\a}{}^{\b} \nabla^a$ refers to a vector covariant derivative in bi-spinor representation\footnote{Note that the vector covariant derivatives above should also be projected to components as they contain implicit fermionic component terms not given here. These terms are found by degauging the derivatives which is the topic of Subsection \ref{sec:cadabraDegauge}}. While logically straightforward, this is a time-consuming task for invariants constructed from more complicated composite fields than that of Eq. (\ref{eq:compositeCadabra}) and may require nontrivial uses of properties and symmetries to recognize analytically equivalent structures. In fact, as we touch on in the next section, to analyze a full linearly independent set of curvature-squared invariants in components for $5D, \cN =1$ supergravity, the tedium and complexity in the component reduction process approaches a scale that is very likely impossible without computational aid. This is where \cadabrainline{cadabra} comes in.

Let us proceed by demonstrating a simple example in which we apply spinor covariant derivatives on the Weyl multiplet superfield and one of its descendants. First we must declare the necessary properties and objects in code including indices, a Kronecker delta, field symmetries, and covariant derivatives. Also, we define the ``tower" operations of Eqs. (\ref{cadabraWdervs-a}). 
%

%\noindent\line(1,0){\textwidth}
\noindent\rule{\textwidth}{0.4pt}
\vspace{-2em}
\begin{cadabra}
## Indices ##
{\a, \b, \p, \g, \alpha#}::Indices(spinor, position=independent).
{\a, \b, \p, \g, \alpha#}::Integer(1..4).
{i, j, k, l, i#}::Indices(su2, position=independent).
{i, j, k, l, i#}::Integer(1..2).
{A, B}::Indices(DummySpinorsu2, position=independent).
{\A, \B}::Indices(DummySpinorSpinor, position=independent).

## Field and Object Properties ##
\d{#}::KroneckerDelta.
\D{#}::Derivative. ## Spinor covariant derivative.
\vD{#}::Derivative. ## Vector derivative.
W_{\a \b}::Symmetric.
\fW_{\a \b \g}^{k}::TableauSymmetry(shape={3}, indices={0,1,2}).
\wX_{\a \b}^{i j}::TableauSymmetry(shape={2}, indices={0,1}, 
                                   shape={2}, indices={2,3}).
{W{#}, \fW{#}, \fX{#}}::Depends(\D{#}, \vD{#}).

## Superspace Expressions ##
component_reduce_W := \Lambda^{i}_{\g}^{A} \D_{A}{W_{\a \b}};
component_reduce_X := \Lambda^{i}_{\a}^{A} \D_{A}{\fX^{j}_{\b}};
\end{cadabra}
\begin{flalign}
    ~~~~~&(\Lambda)^i_\g{}^A \nabla_A W_{\a \b} && \nonumber \\
    ~~~~~&(\Lambda)^i_\a{}^A \nabla_A X^j_\b && \nonumber
\end{flalign}
\noindent\rule{\textwidth}{0.4pt}
Note that semicolons instruct \cadabrainline{cadabra} notebooks to display in rendered \cadabrainline{LaTeX}, and some \cadabrainline{LaTeX} commands above do not exist by default. \cadabrainline{Cadabra} properties not displayed here are used to define these macros, i.e., \cadabrainline{\\a::LaTeXForm("\\alpha")}. Also note the use of a dummy index $A$ encoding one spinor and one SU(2) index. This is a required concession as \cadabrainline{cadabra} does not support multi-indexed derivatives. A dummy index is also needed for the bi-spinor representation of a vector covariant derivative.
\begin{subequations}
\begin{align}
    \nabla^i_\a := (\Lambda)^i_\a{}^{A} \nabla_A,& ~ ~ (\Lambda)^i_\a{}^{A} (\Lambda^{-1})^{\b}_{j}{}_A = \d^i_j \d^{\b}_\a,\\
    \nabla_{\a \b} := (\Omega)_{\a \b}{}^{\cA} \nabla_{\cA},& ~~ (\Omega)_{\a \b}{}^{\cA} (\Omega^{-1})^{\r \l}{}_{\cA} = \d_{(\a}^{\r} \d_{\b)}^{\l}~.
\end{align}
\end{subequations}
Now to find the above expression in components, we rely heavily on the \cadabrainline{substitute()} core \cadabrainline{cadabra} algorithm. When using this with the substitution rules defined below, the superspace expressions will automatically be replaced with their component-reduced expressions and then can be further reduced by other core algorithms.

\noindent\rule{\textwidth}{0.4pt}
\vspace{-2em}
\begin{cadabra}
## Substitution Rules ##
tower_project_weyl := {
  \D_{A}{W_{\a \b}} ->  
            \ILambda^{\g}_{k}_{A} \fW_{\a \b \g}^{k} 
    + 1/2   \ILambda^{\g}_{k}_{A} \e_{\g \a} \fX^{k}_{\b}
    + 1/2   \ILambda^{\g}_{k}_{A} \e_{\g \b} \fX^{k}_{\a},
  \D_{A}{\fX^{j}_{\b}} ->   
            \ILambda^{\a}_{i}_{A} \wX_{\a \b}^{i j} 
    + 1/8 I \ILambda^{\a}_{i}_{A} \e^{i j} \e_{\a \b} Y 
    - 3/2 I \ILambda^{\a}_{i}_{A} \e^{i j} \e_{\a \l} 
            \AOmega^{\l \p}^{\A} \vD_{\A}(W_{\b \p}) 
    +   2 I \ILambda^{\a}_{i}_{A} \e^{i j} \e^{\l \p} 
             W_{\a \l} W_{\b \p}
    + 1/2 I \ILambda^{\a}_{i}_{A} \e^{i j} \e_{\a \b} 
            \e^{\g \l} \e^{\p \t} W_{\g \p} W_{\l \t} 
    - 1/2 I \ILambda^{\a}_{i}_{A} \e^{i j} \e_{\b \l} 
            \AOmega^{\l \p}^{\A} \vD_{\A}(W_{\a \p})
}.

remove_lambda := {
  \Lambda^{i}_{\a}^{A} \ILambda^{\b}_{j}_{A} -> \d^{\b}_{\a} \d^{i}_{j}
}.
\end{cadabra}
\begin{cadabra}
component_reduce_W;
substitute(component_reduce_W, tower_project_weyl);
distribute(component_reduce_W)
substitute(component_reduce_W, remove_lambda)
eliminate_kronecker(component_reduce_W);
\end{cadabra}
\begin{flalign}
     ~~~~~& (\Lambda)^{i}_{\g}{}^{A} \nabla_{A}{W_{\a \b}} && \nonumber \\
     ~~~~~& (\Lambda)^i_\g{}^A\left((\Lambda^{-1})^\r_k{}_A W_{\a \b}{}_\r^k + \frac{1}{2} (\Lambda^{-1})^\r_k{}_A \e_{\r \a} X^k_\b + \frac{1}{2} (\Lambda^{-1})^\r_k{}_A \e_{\r \b} X^k_\a \right) && \nonumber \\
     ~~~~~& W_{\a \b}{}^i_\g + \frac{1}{2} \e_{\g \a} X^i_\b + \frac{1}{2} \e_{\g \b} X^i_\a && \nonumber
\end{flalign}
\vspace{-1.5em}
\begin{cadabra}
component_reduce_X;
distribute(component_reduce_X)
unwrap(component_reduce_X)
substitute(component_reduce_X, tower_project_weyl)
distribute(component_reduce_X)
substitute(component_reduce_X, remove_Lambda)
eliminate_kronecker(component_reduce_X);
canonicalise(component_reduce_X);
\end{cadabra}
\begin{flalign}
     ~~~~~&(\Lambda)^i_\a{}^A \nabla_A X^j_\b && \nonumber \\
     ~~~~~& (\Lambda)^i_\a{}^A\bigg((\Lambda^{-1})^{\a_1}_k{}_A X_{{\alpha_1} {\beta}}\,^{k j} + \frac{1}{8} I (\Lambda^{-1})^{\a_1}_k{}_A \eps^{k j} \eps_{{\alpha_1} {\beta}} Y  && \nonumber \\
     ~~~~~&~~~ - \frac{3}{2} I (\Lambda^{-1})^{\a_1}_k{}_A \eps^{k j} \e_{\a_1 \l} (\Omega)^{\l \r}{}^{\mathcal{A}} {\nabla}_\cA {W_{{\beta} {\rho}}} + 2 I (\Lambda^{-1})^{\a_1}_k{}_A \eps^{k j} \e^{\l \r} W_{{\alpha_1\l}} W_{{\beta} {\r}}  && \nonumber \\
     ~~~~~&~~~+\frac{1}{2} I (\Lambda^{-1})^{\a_1}_k{}_A \eps^{k j} \eps_{{\alpha_1} {\beta}} W^{{\gamma} {\delta}} W_{{\gamma} {\delta}}-\frac{1}{2} I (\Lambda^{-1})^{\a_1}_k{}_A \eps^{k j}\e_{\b \l}  (\Omega)^{\l \r}{}^{\cA}{\nabla}_\cA {W_{{\alpha_1} {\rho}}} \bigg) && \nonumber \\
     ~~~~~&X_{{\alpha} {\beta}}\,^{i j} + \frac{1}{8} I \eps^{i j} \eps_{{\alpha} {\beta}} Y - \frac{3}{2} I\eps^{i j} \e_{\a \l} (\Omega)^{\l \r}{}^{\mathcal{A}} {\nabla}_\cA {W_{{\beta} {\rho}}} + 2 I \eps^{i j} \e^{\l \r} W_{{\alpha \l}} W_{{\beta} {\r}}   && \nonumber \\
     ~~~~~&~~~ +\frac{1}{2} I \eps^{i j} \eps_{{\alpha} {\beta}} W^{{\gamma} {\delta}} W_{{\gamma} {\delta}} -\frac{1}{2} I \eps^{i j}\e_{\b \l}  (\Omega)^{\l \r}{}^{\cA}{\nabla}_\cA {W_{{\alpha} {\rho}}}  && \nonumber
\end{flalign}
\noindent\rule{\textwidth}{0.4pt}

Notice the above results indeed match the previously given definitions of Eqs. (\ref{cadabraWdervs-a}) prior to contraction of the antisymmetric tensors and dummy indices and also with symmetries expanded. Explicit antisymmetric tensors are the best way to define object contractions in code as they allow us to write all objects in a canonical form allowing consistent combinations of equivalent structures. They can later be contracted by a custom algorithm for display purposes. This custom algorithm along with various other substitutions including the other tower projections, field properties, generator algebras, and numerous other reduction properties can be found in our repository \cite{GoldGithub} as described in Section \ref{sec:cadabraRepo}.

\section{Optimization} \label{sec:cadabraOptimization}

One of the primary challenges in component reduction by computer algebra is optimization. For example, consider the off-shell ``log'' invariant first found in \cite{Butter:2014xxa} in superspace that is composite of various spinor covariant derivatives acting on a vector multiplet superfield, $W$. The importance of this invariant is that it leads to a supersymmetric completion of the Ricci tensor squared. Its component form was found in \cite{ Gold:2023ykx, Gold:2023dfe} with \cadabrainline{cadabra} using the tower projection rules of the vector multiplet and Weyl multiplet.
\begin{equation}
    G^{i j} \rightarrow H^{i j}_{\textrm{log}} = \frac{3 \ri}{1280} \nabla^{(i j} \nabla^{k l)} \nabla_{k l} \textrm{log} W, \; \; \; \; \; \; \nabla^{i j} := \nabla^{\a (i} \nabla^{j)}_{\a}~.
\end{equation}
Upon examination of this object, it quickly becomes clear that it is an extremely involved task to calculate six spinor covariant derivatives on $\log W$.  Indeed, one must then also generate its descendants by further actions of two spinor covariant derivatives as can be seen in Eqs. (\ref{eq:descCadabra}). Table \ref{table:cadabra} shows the nonlinear increase in complexity of component expressions with increasing spinor covariant derivatives by an optimized approach. When not optimized, the computation is intractable.
%
% \begin{table}[ht]
% \caption{The size of component expressions on the logarithm of a vector multiplet, $W$ increasing in spinor covariant derivatives. The table also includes the maximum number of computationally parsed terms and the elapsed time on a typical personal computer by an optimized algorithm to demonstrate the nonlinear increase in complexity with number of derivatives.}
% \begin{center}
% \begin{tabular}{ |c | c | c | c |}
% \hline
%  Increasing Derivatives & Component Form & Parsed Terms & Time Elapsed \\ \hline\hline
% $ \nabla_{\a l} \textrm{log} W$ & $\ri \lambda_{i \a} W^{-1}$  & 1 & 0.07 s \\ \hline 
% $  \nabla^{\a}_{k} \nabla_{l \a} \textrm{log} W$ & $- 4 \ri X_{k l} W^{-1} + \lambda_{k} \lambda_{l} W^{-2}$ &  5 & 0.14 s \\ \hline
% $\nabla^{i j}\nabla_{k l} \textrm{log} W$ & (56 terms) &  290 & 0.94 s  \\ \hline
% $ \nabla^{(i j} \nabla^{k l)}\nabla_{k l} \textrm{log} W$ & (112 terms) &  20056 &  4.01 min \\ \hline %240.49s
% $\nabla_{i \a} \nabla_{j \b} \nabla^{(i j} \nabla^{k l)}\nabla_{k l} \textrm{log} W$ & (1641 terms) & 87855 & 2.87 hr \\ \hline %10318.09s
% \end{tabular}
% \end{center}
% \label{table:cadabra}
% \end{table}

\begin{table}
\caption{The size of component expressions on the logarithm of a vector multiplet, $W$ increasing in spinor covariant derivatives. The table also includes the maximum number of computationally parsed terms and the elapsed time on a typical personal computer by an optimized algorithm to demonstrate the nonlinear increase in complexity with number of derivatives.}
\label{tab:1}       % Give a unique label
\begin{tabular}{p{3.5cm}p{3.5cm}p{2cm}p{2cm}}
\hline\noalign{\smallskip}
Increasing Derivatives & Component Form & Parsed Terms & Time Elapsed \\
\noalign{\smallskip}\svhline\noalign{\smallskip}
$ \nabla_{\a l} \textrm{log} W$ & $\ri \lambda_{i \a} W^{-1}$  & 1 & 0.07 s \\ 
$  \nabla^{\a}_{k} \nabla_{l \a} \textrm{log} W$ & $- 4 \ri X_{k l} W^{-1} + \lambda_{k} \lambda_{l} W^{-2}$ &  5 & 0.14 s \\ 
$\nabla^{i j}\nabla_{k l} \textrm{log} W$ & (56 terms) &  290 & 0.94 s  \\ 
$ \nabla^{(i j} \nabla^{k l)}\nabla_{k l} \textrm{log} W$ & (112 terms) &  20056 &  4.01 min \\  %240.49s
$\nabla_{i \a} \nabla_{j \b} \nabla^{(i j} \nabla^{k l)}\nabla_{k l} \textrm{log} W$ & (1641 terms) & 87855 & 2.87 hr \\ 
\noalign{\smallskip}\hline\noalign{\smallskip}
\end{tabular}
\label{table:cadabra}
\end{table}
As this table suggests, an optimized approach parses nearly one-hundred thousand terms. As described in \cite{Gold:2023dfe,Gold:2023ykx},  it is also desirable to find descendants of equation of motion multiplets from the resulting BF component action of this invariant that is analogous in complexity to nine or ten spinor covariant derivatives on $\log W$. This means \cadabrainline{cadabra} should be prepared to parse expressions on the order of millions of terms.

Typically when one reduces expressions in \cadabrainline{cadabra}, core algorithms are chosen in order by the examination of an expression's current structure as in Section \ref{sec:cadabraReduction}. Automation is needed for larger expressions, so a post-process method is created to run all necessary reduction algorithms repeatedly until the expression converges to a minimal number of terms. An expression with the above scale will need automation, but while \cadabrainline{cadabra} is fast, it is not fast enough to run an arbitrary number of algorithms on this scale with the complex structures found in conformal supergravity. An automated optimization method is therefore needed to examine existing structures and intelligently run algorithms in order of priority.

To achieve this, we introduce the \pythoninline{python} method below which examines an expression by utilizing the \pythoninline{Ex} and \pythoninline{ExNode} classes. These classes are the fundamental building blocks of expression objects in the underlying \cadabrainline{cadabra} core written in \cadabrainline{C++}.
\begin{python}
from cadabra2 import *

def check_expression_for_object(expression, logical_structure, obj,
                     nested_objects = [], add_parent_object = False):
    for occurance in expression[obj]:
        if nested_objects:
            for child in occurance.children():
                for nested_object in nested_objects:
                    if str(child.name) == nested_object:
                        logical_structure[nested_object] +=1
        if not nested_objects or add_parent_object:
                logical_structure[obj] +=1
    return(logical_structure)
        
\end{python}
Let us continue with an example in code needing automated component reduction so that we can demonstrate this method in use. For this example we define a logical structure or dictionary that contains information on the count of spinor covariant derivatives needing reduction as demonstrated in Section \ref{sec:cadabraReduction}. The structure also contains information on the count of sums, products, powers, and vector covariant derivatives nested inside of spinor covariant derivatives that will need different operations before the tower projection can be performed. Note that \cadabrainline{sum}, \cadabrainline{prod}, and \cadabrainline{pow} are all names of \pythoninline{ExNode} types.

\noindent\rule{\textwidth}{0.4pt}
\vspace{-2em}
\begin{cadabra}
expression := \D_{A}{\cH^{a} W**(-2) + 2 \vD^{a}{\vW}};
\end{cadabra}
\begin{flalign}
    ~~~~~ & \nabla_A\left( \cH^a W^{-2} + 2 \nabla^a W\right) \nonumber &&
\end{flalign}
\vspace{-1.5em}
\begin{cadabra}
logical_structure = {'\\D':0, '\\vD':0, '\\sum':0, '\\prod':0, '\\pow':0}
obj_check = '\\D'
nested_objects = ['\\vD', '\\prod', '\\sum', '\\pow']
check_expression_for_object(expression, logical_structure, 
    obj_check, nested_objects, True);
\end{cadabra}
\begin{flalign}
    ~~~~~& \textrm{\{'\textbackslash \textbackslash D':1, '\textbackslash \textbackslash vD':0, '\textbackslash \textbackslash sum':1, '\textbackslash \textbackslash prod':0, '\textbackslash \textbackslash pow':0\}} && \nonumber
\end{flalign}
\noindent\rule{\textwidth}{0.4pt}
As expected, this has detected a single spinor covariant derivative, but it also has detected a nested sum. We therefore know to run a prerequisite operation \cadabrainline{distribute()} to distribute the sum before attempting to project to components. We repeat this process a few times in Table \ref{table:cadabra2} to show the operations determined by a previous logical structure, its output, and the resulting logical structure that determines the next iteration.
\begin{table}[ht]
\caption{This table gives three iterations of the process where suitable operations are chosen based on a previous logical structure, its output is shown, and a new logical structure is generated. According to the logical structure result of the final iteration, we can deduce a vector derivative is nested inside of a spinor covariant derivative. Therefore we must run vector algebra operation next to move it inside.}
\begin{center}
\begin{tabular}{  l  c }
\hline
~~~~~~~~~~~~ Operations and Output & Resulting Logical Structure \\ \svhline
% Column 1
\begin{cadabra}

distribute(expression)
unwrap(expression);
\end{cadabra}

& $~~${\scriptsize \{'\textbackslash \textbackslash D':2, '\textbackslash \textbackslash vD':1, '\textbackslash \textbackslash sum':0, '\textbackslash \textbackslash prod':1, '\textbackslash \textbackslash pow':0\}}$~~$ \\ 
& \\
{\small $~~~~\nabla_A\left( \cH^a W^{-2} \right)+ 2 \nabla_A \nabla^a W$} & \\ 
& \\
\hline
% Column 2a
\begin{cadabra} 

product_rule(expression)
distribute(expression); 
\end{cadabra}

& $~~${\scriptsize \{'\textbackslash \textbackslash D':3, '\textbackslash \textbackslash vD':1, '\textbackslash \textbackslash sum':0, '\textbackslash \textbackslash prod':0, '\textbackslash \textbackslash pow':1\}}$~~$ \\ 
% Column 2b
& \\
{\small $~~~~\nabla_A \cH^a W^{-2} + \cH^a \nabla_A W^{-2} + 2 \nabla_A \nabla^a W$} & \\ 
% Column 2b
& \\
\hline
% Column 3
\begin{cadabra}

product_rule(expression)
distribute(expression);
\end{cadabra}

& $~~${\scriptsize \{'\textbackslash \textbackslash vD':1, '\textbackslash \textbackslash D':3, '\textbackslash \textbackslash sum':0, '\textbackslash \textbackslash prod':0, '\textbackslash \textbackslash pow':0\}}$~~$ \\
% Column 2b
& \\
{\small $~~~~\nabla_A \cH^a W^{-2} -2 \cH^a W^{-3}\nabla_A W + 2 \nabla_A \nabla^a W$} ~& \\ 
& \\
\hline
\end{tabular}
\end{center}
\label{table:cadabra2}
\end{table}
While this process may seem convoluted, remember that in any useful \cadabrainline{cadabra} computation the expressions are so large that a manual check of the next optimal operation is time consuming at best and impossible at worst. With this method and other methods like it, we determine a set of rules to automatically run the appropriate order of operations for each resulting logical structure. These are all contained in a class named \cadabrainline{CleverSimplify} that automatically reduces superspace expressions needing component reduction, and it is also capable of performing other reduction procedures that are discussed in the next section. Below we generate up to four derivatives on $\log W$ to demonstrate it in use.

\noindent\rule{\textwidth}{0.4pt}
\vspace{-2em}
\begin{cadabra}
## Two spinor covariant derivatives on logW. ##
## CleverSimplify will stop when finished or at 15 iterations.
maximum_iterations = 15
twoDerivatives := \e^{\a \b} \e_{k i} \e_{l j} \Lambda_{k}^{\a}^{A}  
                  \Lambda_{l \a}^{B} \D_{A}{\D_{B}{\log(\vW)}};
substitute(twoDerivatives, $\D_{A}{\log{X?}} -> X?**(-1) \D_{A}{X?}$);
componentReduce = CleverSimplify(twoDerivatives)
componentReduce.main(maximum_iterations);
\end{cadabra}
%\vspace{-1em}
{\small
\begin{flalign}
    ~~~~~~&\e^{\a \b}\e_{k i} \e_{l j} (\Lambda)^i_\b{}^A (\Lambda)^j_\a{}^B \nabla_A \nabla_B \left( \log (W)\right) && \nonumber \\
    ~~~~~~&\e^{\a \b}\e_{k i} \e_{l j} (\Lambda)^i_\b{}^A (\Lambda)^j_\a{}^B \nabla_A \left( W^{-1} \nabla_B W\right) && \nonumber \\
    ~~~~~~&- 4 I \e_{k i}\e_{l j} W^{-1} X^{i j} - \e^{\a \b} \e_{k i} \e_{l j} W^{-2} \l^i_\a \l^j_\b && \nonumber
\end{flalign}
}
\vspace{-1.5em}
%\vspace{-2.5em}
\begin{cadabra}
## Four spinor covariant derivatives on logW. ##
## CleverSimplify will stop when finished or at 20 iterations.
maximum_iterations = 20 
fourDerivatives :=  \e^{\a \b}\Lambda^{i}_{\b}^{A} \Lambda^{j}_{\a}^{B}  
                    \D_{A}{\D_{B}{@(twoDerivatives)}};
componentReduce = CleverSimplify(fourDerivatives)
componentReduce.main(maximum_iterations);
\end{cadabra}
{\small
\begin{flalign}
&\epsilon^{\alpha \beta} (\Lambda)^{i}_{\beta}{}^{A} (\Lambda)^{j}_{\alpha}{}^{B} \nabla_{A}{\nabla_{B}({-4I \epsilon_{k {i_{1}}} \epsilon_{l {i_{2}}} {W}^{-1} X^{{i_{1}} {i_{2}}}} -\epsilon^{\rho \gamma} \epsilon_{k i_1} \epsilon_{l i_2} {W}^{-2} \lambda^{{i_{1}}}_{\rho} \lambda^{{i_{2}}}_{\gamma}}) \nonumber \\
&{}-8I \epsilon^{\alpha \beta} \epsilon_{k {i_{1}}} \epsilon_{l {i_{2}}} {W}^{-3} X^{{i_{1}} {i_{2}}} \lambda^{i}_{\alpha} \lambda^{j}_{\beta}+16\epsilon_{k {i_{1}}} \epsilon_{l {i_{2}}} {W}^{-2} X^{i j} X^{{i_{1}} {i_{2}}}-4I (\tilde{\Omega})^{\alpha \beta \mathcal{A}} \delta^{i}\,_{k} \epsilon_{l {i_{1}}} {W}^{-2} \lambda^{j}_{\alpha} \nabla_{\mathcal{A}}{\lambda^{{i_{1}}}_{\beta}} \nonumber\\
&-4I (\tilde{\Omega})^{\alpha \beta \mathcal{A}} \delta^{i}\,_{l} \epsilon_{k {i_{1}}} {W}^{-2} \lambda^{j}_{\alpha} \nabla_{\mathcal{A}}{\lambda^{{i_{1}}}_{\beta}}+\frac{3}{2}I \delta^{i}\,_{k} \epsilon^{\alpha \beta} \epsilon^{\gamma \rho} \epsilon_{l {i_{1}}} {W}^{-2} W_{\alpha \gamma} \lambda^{j}_{\beta} \lambda^{{i_{1}}}_{\rho}+\frac{3}{2}I \delta^{i}\,_{l} \epsilon^{\alpha \beta} \epsilon^{\gamma \rho} \epsilon_{k {i_{1}}} {W}^{-2} W_{\alpha \gamma} \lambda^{j}_{\beta} \lambda^{{i_{1}}}_{\rho} \nonumber\\
&+3\delta^{i}\,_{k} \epsilon^{\alpha \beta} \epsilon_{l {i_{1}}} {W}^{-1} X^{{i_{1}}}_{\alpha} \lambda^{j}_{\beta}+3\delta^{i}\,_{l} \epsilon^{\alpha \beta} \epsilon_{k {i_{1}}} {W}^{-1} X^{{i_{1}}}_{\alpha} \lambda^{j}_{\beta}-4I (\tilde{\Omega})^{\alpha \beta \mathcal{A}} \delta^{j}\,_{k} \epsilon_{l {i_{1}}} {W}^{-2} \lambda^{i}_{\alpha} \nabla_{\mathcal{A}}{\lambda^{{i_{1}}}_{\beta}} \nonumber \\
&-4I (\tilde{\Omega})^{\alpha \beta \mathcal{A}} \delta^{j}\,_{l} \epsilon_{k {i_{1}}} {W}^{-2} \lambda^{i}_{\alpha} \nabla_{\mathcal{A}}{\lambda^{{i_{1}}}_{\beta}} - \frac{15}{2}I \delta^{j}\,_{k} \epsilon^{\alpha \beta} \epsilon^{\gamma \rho} \epsilon_{l {i_{1}}} {W}^{-2} W_{\alpha \gamma} \lambda^{i}_{\beta} \lambda^{{i_{1}}}_{\rho} - \frac{15}{2}I \delta^{j}\,_{l} \epsilon^{\alpha \beta} \epsilon^{\gamma \rho} \epsilon_{k {i_{1}}} {W}^{-2} W_{\alpha \gamma} \lambda^{i}_{\beta} \lambda^{{i_{1}}}_{\rho} \nonumber \\
&- \frac{15}{2}\delta^{j}\,_{l} \epsilon^{\alpha \beta} \epsilon_{k {i_{1}}} {W}^{-1} X^{{i_{1}}}_{\alpha} \lambda^{i}_{\beta}+4(\tilde{\Omega})^{\alpha \beta \mathcal{A}} (\tilde{\Omega})^{\gamma \rho \mathcal{B}} \delta^{i}\,_{l} \delta^{j}\,_{k} \epsilon_{\alpha \gamma} \epsilon_{\beta \rho} {W}^{-1} \nabla_{\mathcal{A}}{\nabla_{\mathcal{B}}{W}}+24\delta^{i}\,_{l} \delta^{j}\,_{k} \epsilon^{\alpha \beta} \epsilon^{\gamma \rho} {W}^{-1} W_{\alpha \gamma} F_{\beta \rho} \nonumber \\
&+27\delta^{i}\,_{l} \delta^{j}\,_{k} \epsilon^{\alpha \beta} \epsilon^{\gamma \rho} W_{\alpha \gamma} W_{\beta \rho} - \frac{15}{4}\delta^{j}\,_{k} \epsilon^{\alpha \beta} \epsilon_{l {i_{1}}} {W}^{-1} X^{i}_{\alpha} \lambda^{{i_{1}}}_{\beta}+\frac{15}{2}\delta^{i}\,_{l} \delta^{j}\,_{k} \epsilon^{\alpha \beta} \epsilon_{{i_{1}} {i_{2}}} {W}^{-1} X^{{i_{1}}}_{\alpha} \lambda^{{i_{2}}}_{\beta} \nonumber \\
&- \frac{15}{2}\delta^{j}\,_{k} \epsilon^{\alpha \beta} \epsilon_{l {i_{1}}} {W}^{-1} X^{{i_{1}}}_{\alpha} \lambda^{i}_{\beta}%
+4(\tilde{\Omega})^{\alpha \beta \mathcal{A}} (\tilde{\Omega})^{\gamma \rho \mathcal{B}} \delta^{i}\,_{k} \delta^{j}\,_{l} \epsilon_{\alpha \gamma} \epsilon_{\beta \rho} {W}^{-1} \nabla_{\mathcal{A}}{\nabla_{\mathcal{B}}{W}}+24\delta^{i}\,_{k} \delta^{j}\,_{l} \epsilon^{\alpha \beta} \epsilon^{\gamma \rho} {W}^{-1} W_{\alpha \gamma} F_{\beta \rho} \nonumber \\
&+27\delta^{i}\,_{k} \delta^{j}\,_{l} \epsilon^{\alpha \beta} \epsilon^{\gamma \rho} W_{\alpha \gamma} W_{\beta \rho} - \frac{15}{4}\delta^{j}\,_{l} \epsilon^{\alpha \beta} \epsilon_{k {i_{1}}} {W}^{-1} X^{i}_{\alpha} \lambda^{{i_{1}}}_{\beta}+\frac{15}{2}\delta^{i}\,_{k} \delta^{j}\,_{l} \epsilon^{\alpha \beta} \epsilon_{{i_{1}} {i_{2}}} {W}^{-1} X^{{i_{1}}}_{\alpha} \lambda^{{i_{2}}}_{\beta} - \frac{3}{2}\delta^{i}\,_{l} \delta^{j}\,_{k} Y \nonumber \\
&- \frac{3}{2}\delta^{i}\,_{k} \delta^{j}\,_{l} Y-6\epsilon^{\alpha \beta} \epsilon^{\gamma \rho} \epsilon_{k {i_{1}}} \epsilon_{l {i_{2}}} {W}^{-4} \lambda^{i}_{\alpha} \lambda^{j}_{\beta} \lambda^{{i_{1}}}_{\gamma} \lambda^{{i_{2}}}_{\rho}-8I \epsilon^{\alpha \beta} \epsilon_{k {i_{1}}} \epsilon_{l {i_{2}}} {W}^{-3} X^{i j} \lambda^{{i_{1}}}_{\alpha} \lambda^{{i_{2}}}_{\beta} \nonumber \\
&+4I \delta^{i}\,_{k} \epsilon^{\alpha \beta} \epsilon^{\gamma \rho} \epsilon_{l {i_{1}}} {W}^{-3} F_{\alpha \gamma} \lambda^{j}_{\beta} \lambda^{{i_{1}}}_{\rho}+2I \epsilon^{\alpha \beta} \epsilon_{k {i_{1}}} \epsilon_{l {i_{2}}} {W}^{-3} X^{i {i_{1}}} \lambda^{j}_{\alpha} \lambda^{{i_{2}}}_{\beta}+2I (\tilde{\Omega})^{\alpha \beta \mathcal{A}} \delta^{i}\,_{k} \epsilon_{l {i_{1}}} {W}^{-3} \lambda^{j}_{\alpha} \lambda^{{i_{1}}}_{\beta} \nabla_{\mathcal{A}}{W} \nonumber \\
&+4I \delta^{i}\,_{l} \epsilon^{\alpha \beta} \epsilon^{\gamma \rho} \epsilon_{k {i_{1}}} {W}^{-3} F_{\alpha \gamma} \lambda^{j}_{\beta} \lambda^{{i_{1}}}_{\rho}+2I \epsilon^{\alpha \beta} \epsilon_{k {i_{1}}} \epsilon_{l {i_{2}}} {W}^{-3} X^{i {i_{2}}} \lambda^{j}_{\alpha} \lambda^{{i_{1}}}_{\beta}+2I (\tilde{\Omega})^{\alpha \beta \mathcal{A}} \delta^{i}\,_{l} \epsilon_{k {i_{1}}} {W}^{-3} \lambda^{j}_{\alpha} \lambda^{{i_{1}}}_{\beta} \nabla_{\mathcal{A}}{W} \nonumber \\
&+4I \delta^{j}\,_{k} \epsilon^{\alpha \beta} \epsilon^{\gamma \rho} \epsilon_{l {i_{1}}} {W}^{-3} F_{\alpha \gamma} \lambda^{i}_{\beta} \lambda^{{i_{1}}}_{\rho}+2I \epsilon^{\alpha \beta} \epsilon_{k {i_{1}}} \epsilon_{l {i_{2}}} {W}^{-3} X^{j {i_{1}}} \lambda^{i}_{\alpha} \lambda^{{i_{2}}}_{\beta}+2I (\tilde{\Omega})^{\alpha \beta \mathcal{A}} \delta^{j}\,_{k} \epsilon_{l {i_{1}}} {W}^{-3} \lambda^{i}_{\alpha} \lambda^{{i_{1}}}_{\beta} \nabla_{\mathcal{A}}{W} \nonumber \\
&+3I (\tilde{\Omega})^{\alpha \beta \mathcal{A}} \delta^{j}\,_{k} \epsilon_{l {i_{1}}} {W}^{-2} \lambda^{{i_{1}}}_{\alpha} \nabla_{\mathcal{A}}{\lambda^{i}_{\beta}}-I (\tilde{\Omega})^{\alpha \beta \mathcal{A}} \epsilon^{i j} \epsilon_{k {i_{1}}} \epsilon_{l {i_{2}}} {W}^{-2} \lambda^{{i_{2}}}_{\alpha} \nabla_{\mathcal{A}}{\lambda^{{i_{1}}}_{\beta}} +I (\tilde{\Omega})^{\alpha \beta \mathcal{A}} \delta^{i}\,_{k} \epsilon_{l {i_{1}}} {W}^{-2} \lambda^{{i_{1}}}_{\alpha} \nabla_{\mathcal{A}}{\lambda^{j}_{\beta}} \nonumber \\
&+\frac{3}{4}\epsilon^{\alpha \beta} \epsilon^{i j} \epsilon_{k {i_{1}}} \epsilon_{l {i_{2}}} {W}^{-1} X^{{i_{1}}}_{\alpha} \lambda^{{i_{2}}}_{\beta} - \frac{3}{4}\delta^{i}\,_{k} \epsilon^{\alpha \beta} \epsilon_{l {i_{1}}} {W}^{-1} X^{j}_{\alpha} \lambda^{{i_{1}}}_{\beta}-4\delta^{i}\,_{l} \delta^{j}\,_{k} \epsilon^{\alpha \beta} \epsilon^{\gamma \rho} {W}^{-2} F_{\alpha \gamma} F_{\beta \rho} \nonumber \\
&-4\epsilon_{k {i_{1}}} \epsilon_{l {i_{2}}} {W}^{-2} X^{i {i_{2}}} X^{j {i_{1}}}-(\tilde{\Omega})^{\alpha \beta \mathcal{A}} (\tilde{\Omega})^{\gamma \rho \mathcal{B}} \delta^{i}\,_{l} \delta^{j}\,_{k} \epsilon_{\alpha \gamma} \epsilon_{\beta \rho} {W}^{-2} \nabla_{\mathcal{A}}{W} \nabla_{\mathcal{B}}{W}+4I \delta^{j}\,_{l} \epsilon^{\alpha \beta} \epsilon^{\gamma \rho} \epsilon_{k {i_{1}}} {W}^{-3} F_{\alpha \gamma} \lambda^{i}_{\beta} \lambda^{{i_{1}}}_{\rho} \nonumber \\
&+2I \epsilon^{\alpha \beta} \epsilon_{k {i_{1}}} \epsilon_{l {i_{2}}} {W}^{-3} X^{j {i_{2}}} \lambda^{i}_{\alpha} \lambda^{{i_{1}}}_{\beta}+2I (\tilde{\Omega})^{\alpha \beta \mathcal{A}} \delta^{j}\,_{l} \epsilon_{k {i_{1}}} {W}^{-3} \lambda^{i}_{\alpha} \lambda^{{i_{1}}}_{\beta} \nabla_{\mathcal{A}}{W}-4\delta^{i}\,_{k} \delta^{j}\,_{l} \epsilon^{\alpha \beta} \epsilon^{\gamma \rho} {W}^{-2} F_{\alpha \gamma} F_{\beta \rho} \nonumber \\
&-4\epsilon_{k {i_{1}}} \epsilon_{l {i_{2}}} {W}^{-2} X^{i {i_{1}}} X^{j {i_{2}}}-(\tilde{\Omega})^{\alpha \beta \mathcal{A}} (\tilde{\Omega})^{\gamma \rho \mathcal{B}} \delta^{i}\,_{k} \delta^{j}\,_{l} \epsilon_{\alpha \gamma} \epsilon_{\beta \rho} {W}^{-2} \nabla_{\mathcal{A}}{W} \nabla_{\mathcal{B}}{W}+3I (\tilde{\Omega})^{\alpha \beta \mathcal{A}} \delta^{j}\,_{l} \epsilon_{k {i_{1}}} {W}^{-2} \lambda^{{i_{1}}}_{\alpha} \nabla_{\mathcal{A}}{\lambda^{i}_{\beta}} \nonumber \\
&-I (\tilde{\Omega})^{\alpha \beta \mathcal{A}} \epsilon^{i j} \epsilon_{k {i_{1}}} \epsilon_{l {i_{2}}} {W}^{-2} \lambda^{{i_{1}}}_{\alpha} \nabla_{\mathcal{A}}{\lambda^{{i_{2}}}_{\beta}}+I (\tilde{\Omega})^{\alpha \beta \mathcal{A}} \delta^{i}\,_{l} \epsilon_{k {i_{1}}} {W}^{-2} \lambda^{{i_{1}}}_{\alpha} \nabla_{\mathcal{A}}{\lambda^{j}_{\beta}}+\frac{3}{4}\epsilon^{\alpha \beta} \epsilon^{i j} \epsilon_{k {i_{1}}} \epsilon_{l {i_{2}}} {W}^{-1} X^{{i_{2}}}_{\alpha} \lambda^{{i_{1}}}_{\beta} \nonumber \\
&- \frac{3}{4}\delta^{i}\,_{l} \epsilon^{\alpha \beta} \epsilon_{k {i_{1}}} {W}^{-1} X^{j}_{\alpha} \lambda^{{i_{1}}}_{\beta} \nonumber
\end{flalign}
}
\noindent\rule{\textwidth}{0.4pt}

Continuing this process up to six derivatives yields the desired field, $H^{i j}_{\log}$, in components, but it is excessive to give it here. This is a novel result that first appeared in Eq. (3.25) of \cite{Gold:2023dfe}, after some further reduction by hand. Also note that various substitutions and properties not given here are used by \cadabrainline{CleverSimplify} including the remaining tower projections, field properties, and algebras and are given in our repository \cite{GoldGithub}. See Section \ref{sec:cadabraRepo} for more information.

\section{Other Reduction Procedures} \label{sec:cadabraOther}

\subsection{Testing} \label{sec:cadabraTesting}

It is important to ensure accuracy when generating large expressions in code. A generated object of hundreds or thousands of terms is at risk for errors as any minor typo or logical mistake in code can propagate to many terms. There are two primary strategies to ensuring accuracy. First, after major changes in code, it is important to generate known component expressions and match them with those in the literature. These could be considered sanity checks or unit tests, and they range from small identities to complete actions. Second, consistency checks are essential for new component expressions not existing in literature.

For example, the linear multiplet superfield satisfies the following properties by construction.
\bsubeq
\begin{align}
    K^a G^{i j} &= 0 ~, \\
    S^k_\a G^{i j} &= 0 ~, \\
    \nabla^{(k}_\a G^{i j)} &= 0 ~,
\end{align}
\esubeq
where $K^a$ is the generator of special conformal transformations and $S_\a^k$ the generator of $S$-supersymmetry transformations within the local $5D$ $\cN=1$ superconformal algebra \cite{Butter:2014xxa}, and in the last expression we are completely symmetrising SU(2) indices.
All of the previously mentioned composite linear multiplet fields used in the BF action should therefore satisfy these properties in components. See below as \cadabrainline{CleverSimplify} proves this to be true with the composite field $H^{i j}_{\textrm{Weyl}}$ thereby satisfying these consistency checks.

\noindent\rule{\textwidth}{0.4pt}
\vspace{-2em}
\begin{cadabra}
## Definition of HijWeyl and simple consistency checks. ##
consitency_check = True
HijWeyl:= - (I/2) \e^{\l \a} \e^{\p \b} \e^{\t \g} 
                   \fW_{\a \b \g}^{i} \fW_{\l \p \t}^{j} 
          + (3 I / 2) \e^{\a \p} \e^{\b \g} W_{\p \g} \wX_{\a \b}^{i j} 
          - (3 I / 4) \e^{\a \b} \fX^{i}_{\b} \fX^{j}_{\a}; 
\end{cadabra}
\begin{flalign}
~~~~~~& - \frac{1}{2}I \e^{\l \a} \e^{\r \b} \e^{\t \g} W_{\alpha \beta \gamma}\,^{i} W_{\l \r \t}\,^{j}+\frac{3}{2}I \epsilon^{\alpha \r} \epsilon^{\beta \g} W_{\r \g} X_{\alpha \beta}\,^{i j} - \frac{3}{4}I \epsilon^{\alpha \beta} X^{i}_{\beta} X^{j}_{\alpha} && \nonumber
\end{flalign}

\begin{cadabra}
check1 := K^{a}{@(HijWeyl)};
consistencyCheck = CleverSimplify(check1)
consistencyCheck.main(5, consitency_check);

check2 := \Lambda^{k}_{\a}^{A} S_{A}{@(HijWeyl)};
consistencyCheck = CleverSimplify(check2)
consistencyCheck.main(6, consitency_check);

check3 := \Symmetrize_{i j k}^{i1 i2 i3} \Lambda^{k}_{\a}^{A} 
          \D_{A}{@(HijWeyl)};
consistencyCheck = CleverSimplify(check3)
consistencyCheck.main(6, consitency_check);
\end{cadabra}
\begin{flalign}
~~~~~~&{}K^{a}\left({- \frac{1}{2}I \e^{\l \a} \e^{\r \b} \e^{\t \g} W_{\alpha \beta \gamma}\,^{i} W_{\l \r \t}\,^{j}+\frac{3}{2}I \epsilon^{\alpha \r} \epsilon^{\beta \g} W_{\r \g} X_{\alpha \beta}\,^{i j} - \frac{3}{4}I \epsilon^{\alpha \beta} X^{i}_{\beta} X^{j}_{\alpha}}\right) && \nonumber \\
~~~~~~&{}0  \nonumber && \\
~~~~~~&{}(\Lambda)^k_\a{}^A S_{A}\left({ - \frac{1}{2}I \e^{\l \a} \e^{\r \b} \e^{\t \g} W_{\alpha \beta \gamma}\,^{i} W_{\l \r \t}\,^{j}+\frac{3}{2}I \epsilon^{\alpha \r} \epsilon^{\beta \g} W_{\r \g} X_{\alpha \beta}\,^{i j} - \frac{3}{4}I \epsilon^{\alpha \beta} X^{i}_{\beta} X^{j}_{\alpha}}\right) && \nonumber \\
~~~~~~&{}0 && \nonumber \\
~~~~~~&{}\mathbb{S}_{i j k}\,^{{i_{1}} {i_{2}} {i_{3}}} \left( (\Lambda)^{k}_{\alpha}{}^{A} \nabla_{A}({ - \frac{1}{2}I \e^{\l \a} \e^{\r \b} \e^{\t \g} W_{\alpha \beta \gamma}\,^{i} W_{\l \r \t}\,^{j}+\frac{3}{2}I \epsilon^{\alpha \r} \epsilon^{\beta \g} W_{\r \g} X_{\alpha \beta}\,^{i j} - \frac{3}{4}I \epsilon^{\alpha \beta} X^{i}_{\beta} X^{j}_{\alpha}} \right) && \nonumber \\
~~~~~~&{}0 && \nonumber
\end{flalign}
\noindent\rule{\textwidth}{0.4pt}
Note that, for an arbitrary object $O$, we define the following object in code.
\begin{align}
    \mathbb{S}_{i j k}{}^{i_1 i_2 i_3} O^{i j k} := O^{(i_1 i_2 i_3)} ~.
\end{align}
Of course, there are many missing substitution rules regarding symmetry, tower projection, and generators that need to be defined for the above to work and are given in the repository \cite{GoldGithub}. As a final note, all novel results that appear in \cite{Gold:2023dfe, Gold:2023ykx} have undergone various consistency checks of this type some of which are hugely nontrivial checks.

\subsection{Representation Conversion} \label{sec:cadabraRepConvert}
A 5-vector $V^a$ and an antisymmetric tensor $W^{a b}= - W^{b a}$ can be equivalently represented as bi-spinors as 
\bsubeq
\begin{align}
    &V_{\a \b} = V^a (\Gamma_a)_{\a \b}, ~~~ V^a = - \frac{1}{4} (\Gamma_a)^{\a \b}V_{\a \b} ~,\\
    &F_{\a \b} = \frac{1}{2} F^{a b}(\Sigma_{a b})_{\a \b} , ~~~ F_{a b} = (\Sigma_{a b})^{\a \b}F_{\a \b} ~,
\end{align}
\esubeq
where $\Gamma_a$ are the $5D$ gamma matrices and $\Sigma_{a b} = - \frac{1}{4}[\Gamma_a, \Gamma_b]$. It 
is often necessary to convert between these representations to aid in reduction or to proceed with physical analyses. To handle this, simple substitutions are defined, and the \cadabrainline{CleverSimplify} class then automatically reduces expressions as demonstrated in the following example.

\noindent\rule{\textwidth}{0.4pt}
\vspace{-2em}
\begin{cadabra}
spinor_to_vector := {
  W_{\a \b} -> (1/2)\Sigma_{a b}_{\a \b} W^{a b}
}.

vector_to_spinor := {
  W_{a b} -> \Sigma_{a b}_{\p \g} \e^{\a \p} \e^{\b \g} W_{\a \b},
  W^{a b} -> \eta^{a e1} \eta^{b e2} \Sigma_{e1 e2}_{\p \g} 
             \e^{\a \p} \e^{\b \g} W_{\a \b}
}.
\end{cadabra}

\begin{cadabra}
exp := \e^{\a \p} \e^{\b \g} W_{\a \b} W_{\p \g};
substitute(exp, spinor_to_lorentz)
repChange = CleverSimplify(exp)
repChange.main(5);
substitute(exp, lorentz_to_spinor)
repChange = CleverSimplify(exp)
repChange.main(5); 
\end{cadabra}
\begin{flalign}
~~~~~~&\e^{\a \r} \e^{\b \g} W_{\a \b} W_{\r \g} && \nonumber \\
~~~~~~& \frac{1}{2} W_{a b}W^{a b} && \nonumber \\
~~~~~~& \e^{\a \r} \e^{\b \g} W_{\a \b} W_{\r \g} && \nonumber
\end{flalign}
\noindent\rule{\textwidth}{0.4pt}

Field properties, gamma matrix symmetries, and gamma matrix reduction identities are all needed for these reductions and are included in the repository \cite{GoldGithub}.

\subsection{Degauging} \label{sec:cadabraDegauge}
The component vector covariant derivative $\nabla_a$ is defined to coincide with the projection of the superspace derivative $\nabla_a \vert$.
\begin{align} \label{eq:degaugeCad}
    \nabla_a = e_a{}^m \nabla_m = e_a{}^m \left(\mathcal{D}_m - \frac{1}{2} \psi_m{}^\a_i \nabla_\a^i \vert  - \frac{1}{2} \phi_m{}^{\a i} S_{\a i}  - f_m{}^a K_a\right) ~,
\end{align}
where the covariant vector derivative $\mathcal{D}_m$ contains the Lorentz, dilatation, and SU(2) connections. Replacing vector covariant derivatives with this expansion and projecting to components is known as degauging. In the previous sections, we neglected the fact that a complete component expression with fermions requires the degauging of all vector covariant derivatives as in Eq. (\ref{eq:degaugeCad}). Also, for gauge fixing and physical analyses, we may need to degauge other symmetries from the derivative to expose connection fields. For example, the special conformal connection $f_m{}^a$ is composite of Lorentz curvature which is needed to retrieve the Einstein--Hilbert contribution of an action and curvature-squared contributions. Let us demonstrate a bosonic example of how \cadabrainline{CleverSimplify} can be used to determine the $f_{a b} = e_{a}{}^{m} f_{m b}$ contribution in $\Box W \vert$.

\noindent\rule{\textwidth}{0.4pt}
\vspace{-2em}
\begin{cadabra}
spinor_to_lorentz := {
  \vD_{\A}{X??} -> \eta_{a b} \IAOmega_{\a \b}_{\A} 
                   \Gamma^{b}^{\a \b} \vD^{a}{X??}
}.


expose_2Der_spinor := {
  \vD_{\A}{\vD_{\B}{X??}} -> \mD_{\A}{\vD_{\B}{X??}} 
    - \IAOmega_{\a \b}_{\A} \Gamma^{e1}^{\a \b} 
       f_{e1 e2} K^{e2}{\vD_{\B}{X??}}
}.

expose_1Der_spinor := {
  \vD_{\A}{X??} -> \mD_{\A}{X??}  
    - \IAOmega_{\a \b}_{\A} \Gamma^{e1}^{\a \b} f_{e1 e2} K^{e2}{X??}
}.

lorentz_to_spinor := {
  \mD^{a}{X??} -> - (1/4) \Gamma^{a}_{\a \b} \AOmega^{\a \b}^{\A} 
                          \mD_{\A}{X??}
}.
\end{cadabra}

\begin{cadabra}
exp := \eta_{a b} \vD^{a}{\vD^{b}{\vW}};
substitute(exp, lorentz_to_spinor)
unwrap(exp);
repChange = CleverSimplify(exp)
repChange.main(5);
substitute(exp, expose_2Der_spinor)
degauge = CleverSimplify(exp)
degauge.main(5,True)
substitute(exp, expose_1Der_spinor)
degauge = CleverSimplify(exp)
degauge.main(5,True);
substitute(exp, spinor_to_lorentz)
repChange = CleverSimplify(exp)
repChange.main(5,True);
\end{cadabra}
\begin{flalign}
~~~~~~&\eta_{a b} \nabla^a \nabla^b W && \nonumber \\
~~~~~~& - \frac{1}{4}(\tilde{\Omega})^{\a \b \cA}(\tilde{\Omega})^{\g \r \cB} \e_{\a \g} \e_{\b \r} \nabla_\cA \nabla_\cB W  && \nonumber \\
~~~~~~& - \frac{1}{4}(\tilde{\Omega})^{\a \b \cA}(\tilde{\Omega})^{\g \r \cB} \e_{\a \g} \e_{\b \r} \cD_\cA \cD_\cB W - 2 \eta^{a b} W f_{a b}&& \nonumber \\
~~~~~~&\eta_{a b} \cD^a \cD^b W - 2 \eta^{a b} W f_{a b} && \nonumber
\end{flalign}
\noindent\rule{\textwidth}{0.4pt}

The rules for generator and vector derivative algebra, generator actions on fields, new field properties, and more are given in the repository \cite{GoldGithub}.

\section{Repository} \label{sec:cadabraRepo}

\begin{figure}[h]
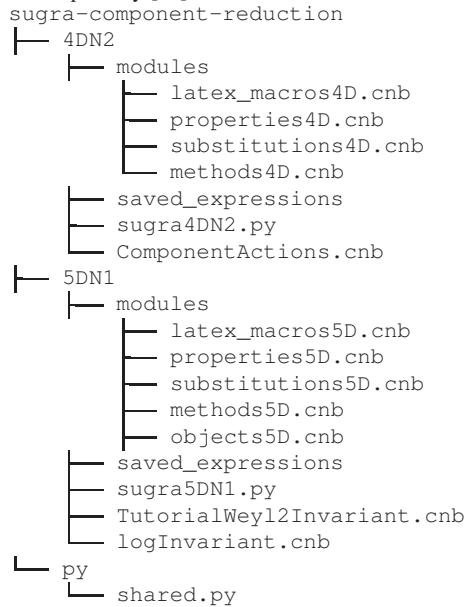

    \footnotesize
    \caption{Structure of the GitHub repository \cite{GoldGithub}.}
    \centering
    \begin{BVerbatim}
    sugra-component-reduction
    ├── 4DN2
        ├── modules
            ├── latex_macros4D.cnb
            ├── properties4D.cnb
            ├── substitutions4D.cnb
            └── methods4D.cnb
        ├── saved_expressions
        ├── sugra4DN2.py
        └── ComponentActions.cnb
    ├── 5DN1
        ├── modules
            ├── latex_macros5D.cnb
            ├── properties5D.cnb
            ├── substitutions5D.cnb
            ├── methods5D.cnb
            ├── objects5D.cnb
        ├── saved_expressions
        ├── sugra5DN1.py
        ├── TutorialWeyl2Invariant.cnb
        └── logInvariant.cnb
    └── py
        └── shared.py
    \end{BVerbatim}
    \label{figure:repoStructure}
\end{figure}

To conclude our paper, we introduce the repository hosted on Github that is packaged in a complete way to perform component reduction on superspace expressions in $4D$, $\mathcal{N}=2$ and $5D$, $\mathcal{N}=1$ matter-coupled supergravity\footnote{We formally ask the reader to cite this paper if using this repository in a public work.}. The main purpose of this section is to provide a guide for the repository's structure (see Figure \ref{figure:repoStructure}) and offer general documentation on essential \pythoninline{python} classes built on top of the \cadabrainline{cadabra} core. It is \textit{strongly} advised to clone the repository locally, after installing \cadabrainline{cadabra}, and use it as described in further detail on the repository page \cite{GoldGithub}. This is ultimately the best way to learn how to use any computational tool, but for completeness, we provide a brief overview here.

Notice that \cadabrainline{.cnb} notebooks are located in either the \cadabrainline{4DN2} and \cadabrainline{5DN1} folders corresponding to each supported supergravity representation. Both representations have various notebook modules that are imported into any primary notebook along with \pythoninline{python} classes contained in \pythoninline{shared.py}. A description of each module is given in Table \ref{table:notebooks} and also on the Github repository's documentation page \cite{GoldGithub}. A convenient importation wrapper is given for each representation, i.e., \pythoninline{sugra4DN2.py} and \pythoninline{sugra5DN1.py}.

The main notebooks provided in the repository include the following. First, in $5D$ we have the \cadabrainline{TutorialWeyl2Invariant.cnb} notebook which includes a brief tutorial demonstrating a few caveats of this tool. It then proceeds to generate the five-dimensional curvature-squared invariant known as the Weyl-squared BF component action from $H^{i j}_{\textrm{Weyl}}$. Next in $5D$ is the \cadabrainline{logInvariant.cnb} notebook which generates the BF component action from $H^{i j}_{\textrm{log}}$. Both of these actions appear in our works \cite{Gold:2023dfe,PRL,Gold:2023ykx}.  Finally, in $4D$ we have the \cadabrainline{ComponentActions.cnb} notebook which generates magnetically and electrically deformed vector, tensor, and hypermultiplet actions that will appear in a future work.
We stress again that the structure of this section is schematic by design providing a sketch of the Github repository content and its functionalities.

\begin{table}[h!]
\caption{Descriptions of module notebooks. Each of these is imported in all primary notebooks in their respective supergravity representation. }
\begin{center}
\begin{tabularx}{\textwidth}{ c | X }
\hline
Module & Description \\ \svhline
% Column 1
\cadabrainline{latex_macros}
& Shorthand LaTeX notations for various objects are defined here. For example, \cadabrainline{\\a::LaTeXForm("\\alpha")} means that \cadabrainline{\\a} will render as $\alpha$ in notebooks. We use these shorthand notations throughout the repository as demonstrated in previous sections. \\ \hline
% Column 2
\cadabrainline{properties}
& \cadabrainline{Cadabra} properties are used to declare indices and objects. For example, this module defines a $5D$ vector index ranging from 0 to 4, a spinor index from 1 to 4, and an SU(2) index from 1 to 2. Additionally, fermionic anticommuting properties are designated for specific indices; if not specified, all are considered bosonic by default. This module also handles the deceleration of field symmetry properties, generators, and other derivatives.\\  \hline
% Column 3
\cadabrainline{substitions}
& This module contains all necessary substitution definitions used in component reduction and consistency checks, including algebra, generators, covariant spinor derivatives acting on fields, contractions, sigma properties, and more.\\  \hline
% Column 4
\cadabrainline{methods}
& Pythonic methods importing all custom classes as well as stand-alone methods used in component reduction are defined here. This is also where the \cadabrainline{CleverSimplify} class is defined that determines the optimized order in which operations should act to simplify and reduce an expression to components.\\  \hline
% Column 5
\cadabrainline{objects}
& This module includes predefined objects such as fields and actions that are used frequently in specific notebooks.\\  \hline
% Column 6
\cadabrainline{shared}
& This module includes python classes that are shared between $4D$ and $5D$. For example, this notably contains a sorting algorithm that determines how to commute or anti-commute two objects to position them correctly within a term, as described by a sort order.\\  \hline
\end{tabularx}
\end{center}
\label{table:notebooks}
\end{table}

\subsection{Class Documentation}

Here we provide general documentation on essential python classes built on top of the \cadabrainline{cadabra} core. Included in the documentation are a class's initialization with arguments, its frequently used public methods with arguments, and an example of it in use. This aligns with general code documentation practices\footnote{For example, see the package documentation on the \cadabrainline{cadabra} website https://cadabra.science/man.html.} and is intended to be used as a reference when directly working in code. All classes are used frequently in component reduction and the other reduction procedures discussed throughout this paper, and they include a custom sorting class that correctly handles fermions\footnote{Basic fermions are supported in \cadabrainline{cadabra}, but fermionic derivatives pose issues requiring a custom sorting solution.}, a class that saves and reads expressions in text, the \cadabrainline{CleverSimplify} class used throughout this paper, and a class that automatically contracts metrics and dummy objects for display purposes.

\subsubsection{Sorting}

Fermions are only partially implemented in \cadabrainline{cadabra}. Fermionic derivatives especially pose issues with the core algorithm \cadabrainline{sort_product()}. A custom sorting algorithm using anti-commuting indices has been developed for the purpose of this repository. It fully handles fermions including derivatives, and it is packaged in the \pythoninline{Sort} class. As demonstrated in the examples below, derivatives automatically move to the right, and sort order is only used on objects with the same number of derivatives.
\newline \vspace{.15cm}

\noindent\fbox{%
    \parbox{\textwidth}{%
        \pythoninline{class shared.Sort(sort_order:list[str], anti_commute:list[str])}
    }%
}

\noindent\fbox{%
    \parbox{\textwidth}{%
        \pythoninline{Sort.sort(exp:Ex)} $\rightarrow$ \pythoninline{sorted_exp:Ex}
    }%
}

\begin{cadabra}
from sugra5DN1 import *

sort_order = ['X','Y']
anti_commute = ['\\a', '\\b', 'A', 'B']
sorter = Sort(sort_order, anti_commute)

exp := \vD^{a}{X_{\b}} Y_{\a};
sorter.sort(exp);
\end{cadabra}
%\vspace{-1cm}
\begin{flalign}
 ~~~~~~&\nabla^{a}{X_{\b}} Y_{\a}  && \nonumber\\
 ~~~~~~&- Y_{\a} \nabla^{a}{X_{\b}} && \nonumber
 \end{flalign}

\begin{cadabra}
exp := \D_{A}{Y^{a}} \D_{B}{X^{b}};
sorter.sort(exp);
\end{cadabra}
\vspace{-.1cm}
\begin{flalign}
 ~~~~~~&\nabla_{A}Y^{a}  \nabla_{B} X^{b} && \nonumber\\
 ~~~~~~&- \nabla_{B} X^{b} \nabla_{A}Y^{a} & \nonumber
 \end{flalign}
 
\noindent\rule{\textwidth}{0.4pt}
\vspace{-1cm}

\subsubsection{Saving}

It is of importance to save expressions after generation and read for later use. This is achieved by basic method that save expressions to a text file and read text files as strings. They are packaged in the \pythoninline{ExpressionSaver} class. The files are automatically saved and read from the \cadabrainline{saved_expressions} folder of a respective representation. 
\newline \vspace{.15cm}

\noindent\fbox{%
    \parbox{\textwidth}{%
        \pythoninline{class shared.ExpressionSaver(file_path:str)}
    }%
}

\noindent\fbox{%
    \parbox{\textwidth}{%
        \pythoninline{ExpressionSaver.save(exp:Ex, file_name:str)}
    }%
}

\noindent\fbox{%
    \parbox{\textwidth}{%
        \pythoninline{ExpressionSaver.read(exp:Ex)} $\rightarrow$ \pythoninline{exp_string:str}
    }%
}

\begin{cadabra}
from sugra5DN1 import *

file_path = '{local_path}/sugra_component_reduction/4DN2/saved_expressions/'
saver = ExpressionSaver(file_path)

exp := X Y.
saver.save(exp, "exp.txt")
read_exp = Ex(saver.read("exp.txt"))
read_exp;
\end{cadabra}
%\vspace{-1cm}
\begin{flalign}
~~~~~~& X Y && \nonumber
\end{flalign}

\noindent\rule{\textwidth}{0.4pt}
\vspace{-1cm}
\subsubsection{CleverSimplify}

\pythoninline{CleverSimplify} is the primary class used for component reduction from superspace and other procedures as demonstrated in the previous sections. It automatically reduces superspace expressions for which it determines an optimal order of operations given a maximum number of iterations. It is also useful in performing other reduction procedures including consistency checks, representation conversion, and derivative degauging. In addition, it contains various properties that are useful for debugging such as terms per iterations, logical structures per iteration, and time elapsed.
\newline \vspace{.15cm}

\noindent\fbox{%
    \parbox{\textwidth}{%
        \pythoninline{class methods.CleverSimlify(exp:Ex)}
    }%
}

\noindent\fbox{%
    \parbox{\textwidth}{%
        \pythoninline{CleverSimplify.main(iterations:int, consistency:bool)} $\rightarrow$ \pythoninline{exp:Ex}
    }%
}

\begin{cadabra}
from sugra5DN1 import *

W_exp := \Lambda^{i}_{\g}^{A} \D_{A}{W_{\a \b}};
componentReduce = CleverSimplify(W_exp)
componentReduce.main(10, False);
componentReduce.terms_per_iteration;
componentReduce.logical_structures;
componentReduce.timer.check_time()
\end{cadabra}
%\vspace{-1cm}
\begin{flalign}
    ~~~~~~&(\Lambda)^i_\g \nabla_A W_{\a \b} && \nonumber \\
    ~~~~~~&W_{\a \b \g}{}^i - \frac{1}{2} \e_{\a \g} X^i_\b - \frac{1}{2} \e_{\b \g} X^i_\a && \nonumber \\
    ~~~~~~& \textrm{[1,1,3,3,3]} && \nonumber \\
    ~~~~~~& \textrm{
        {\scriptsize 
            [\{'\textbackslash \textbackslash D':1, '\textbackslash \textbackslash vD':0, '\textbackslash \textbackslash ILambda':0, '\textbackslash \textbackslash sum':0, '\textbackslash \textbackslash epsilon\_contracts':1, $\cdots$ \},
        }
      } && \nonumber \\
    ~~~~~~& \textrm{
        {\scriptsize 
            \{'\textbackslash \textbackslash D':0, '\textbackslash \textbackslash vD':0, '\textbackslash \textbackslash ILambda':3, '\textbackslash \textbackslash sum':1, '\textbackslash \textbackslash epsilon\_contracts':1, $\cdots$ \},
        }
      } && \nonumber \\
     ~~~~~~& \textrm{
        {\scriptsize 
            \{'\textbackslash \textbackslash D':0, '\textbackslash \textbackslash vD':0, '\textbackslash \textbackslash ILambda':0, '\textbackslash \textbackslash sum':0, '\textbackslash \textbackslash epsilon\_contracts':0, $\cdots$\}, $\cdots$]
        }
      } && \nonumber \\
      ~~~~~~& \textrm{0.05s} && \nonumber
\end{flalign}

% \noindent\fbox{%
%     \parbox{\textwidth}{%
%         \pythoninline{CleverSimplify.check_entire_expression_for_object(logic_struct:dict, obj:str,}  $~~~~~~~~~~~~~~~~~~~$      \pythoninline{nested_objects:list, add_parent_object:bool)} $\rightarrow$ \pythoninline{logic_struct:dict}
%     }%
% }

% \noindent\fbox{%
%     \parbox{\textwidth}{%
%         \pythoninline{CleverSimplify.check_terms(logic_struct:dict, contraction_combos:dict} \pythoninline{        , check_for_factors:bool, check_for_levi:bool)} $\rightarrow$ \pythoninline{cleaned_exp:Ex}
%     }%
% }

% \noindent\fbox{%
%     \parbox{\textwidth}{%
%         \pythoninline{CleverSimplify.expression_check(consistency:bool)} $\rightarrow$ \pythoninline{logic_struct:dict}
%     }%
% }

% \noindent\fbox{%
%     \parbox{\textwidth}{%
%         \pythoninline{CleverSimplify.simplify(logic_struct:dict)} 
%     }%
% }

\noindent\rule{\textwidth}{0.4pt}
\vspace{-1cm}
\subsubsection{CleanUp}

Expressions are often difficult to read due to various in-code requirements such as the use of dummy objects and the definition of canonical forms of objects. By a canonical form, we mean that each object has a predefined index structure wherein contractions are explicit with metrics and antisymmetric tensors. For example, the vector multiplet field strength is defined with lower spinor indices in its bi-spinor representation. We use antisymmetric tensors to place them in the desired position, as given by $F^{\a \b}:= \epsilon^{\a \g} \epsilon^{\b \d} F_{\g \d}$. The \pythoninline{CleanUp} class comes with packaged cleaning methods to make the resulting expression more human-readable. Note however that this should only be used after all operations are finished.
\newline \vspace{.15cm}

\noindent\fbox{%
    \parbox{\textwidth}{%
        \pythoninline{class shared.CleanUp(dummy_object:str,} \\
        $~~~~~~~~~~~~$ \pythoninline{dummy_index_positions:list[parent_rel_t], epsilon_object:str)}
    }%
}

\noindent\fbox{%
    \parbox{\textwidth}{%
        \pythoninline{CleanUp.contract_dummy_auto(exp:Ex, der_object:str)} $\rightarrow$ \pythoninline{cleaned_exp:Ex}
    }%
}

\noindent\fbox{%
    \parbox{\textwidth}{%
        \pythoninline{CleanUp.contract_epsilon_auto(exp:Ex)} $\rightarrow$ \pythoninline{cleaned_exp:Ex}
    }%
}

\begin{cadabra}
from sugra5DN1 import *

epsilon_object = '\\e'
dummy_index_positions = [super, super]
dummy_object = '\\AOmega'
der_object = '\\vD'
cleanup = CleanUp(dummy_object, dummy_index_positions, epsilon_object)


exp := \e_{\a \b} \e^{\t \p} \fl^{i}_{\p} \AOmega^{\b \g}^{\A}\vD_{\A}{\vW} ;
substitute(exp, $\vD_{\A}{X??} -> \vD_{\A \B}{X??}$)
cleanup.contract_dummy_auto(exp, der_object)
cleanup.contract_epsilon_auto(exp)
rename_dummies(exp);
\end{cadabra}
\begin{flalign}
~~~~~~&\e_{\a \b} \e^{\t \r} \l^i_\r (\tilde{\Omega})^{\b \g \cA} \nabla_\cA W && \nonumber \\
~~~~~~& \l^{i \t} \nabla_\a{}^\g W && \nonumber
\end{flalign}
\noindent\rule{\textwidth}{0.4pt}

%%%%%%%%%%%%%%%%%%%%%%%%%%%%%%%%%%%
%%%%%%%%%%%%%%%%%%%%%%%%%%%%%%%%%%%

\begin{acknowledgement}
% If you want to include acknowledgments please use the \verb|acknowledgement| environment -- it will automatically render Springer's preferred layout.
We acknowledge the kind hospitality and financial support extended to us at the MATRIX Program ``New Deformations of Quantum Field and Gravity Theories,'' between 22 Jan and 2 Feb 2024. This work was supported by the Australian Research Council (ARC) Future Fellowship FT180100353, ARC Discovery Project DP240101409, and a Capacity Building Package of the University of Queensland. GG and SK were supported by a postgraduate scholarship at the University of Queensland. 
\end{acknowledgement}
%

%%%%%%%%%%%%%%%%%%%%%%%%%%%%%%%%%%%
%%%%%%%%%%%%%%%%%%%%%%%%%%%%%%%%%%%

\begin{footnotesize}

% \bibitem{SUPERSPACE}
% S.~J.~Gates, M.~T.~Grisaru, M.~Rocek and W.~Siegel,
% ``Superspace Or One Thousand and One Lessons in Supersymmetry,''
% Front. Phys. \textbf{58}, 1-548 (1983)
% [arXiv:hep-th/0108200 [hep-th]].

% \bibitem{WB}
% J.~Wess and J.~Bagger,
% {\it Supersymmetry and Supergravity},
% Princeton University Press, Princeton, 1992.

% \bibitem{BK}
% I.~Buchbinder and S.~M.~Kuzenko.
% \emph{Ideas and methods of supersymmetry and supergravity: 
% Or a walk  through superspace},
% IOP, Bristol (1998).

% \bibitem{Freedman:2012zz}
% D.~Z.~Freedman and A.~Van Proeyen,
% \emph{Supergravity},
% Cambridge Univ. Press, 2012.

% \bibitem{Kuzenko:2022skv}
% S.~M.~Kuzenko, E.~S.~N.~Raptakis and G.~Tartaglino-Mazzucchelli,
% ``Superspace Approaches to $\mathscr {N} = \text{1}$ Supergravity,''
% doi:10.1007/978-981-19-3079-9\_40-1
% [arXiv:2210.17088 [hep-th]].

% \bibitem{Kuzenko:2022ajd}
% S.~M.~Kuzenko, E.~S.~N.~Raptakis and G.~Tartaglino-Mazzucchelli,
% ``Covariant Superspace Approaches to $\mathscr {N}=\text{2}$ Supergravity,''
% doi:10.1007/978-981-19-3079-9\_44-1
% [arXiv:2211.11162 [hep-th]].

% \bibitem{Ozkan:2024euj}
% M.~Ozkan, Y.~Pang and E.~Sezgin,
% ``Higher Derivative Supergravities in Diverse Dimensions,''
% [arXiv:2401.08945 [hep-th]].

% \bibliographystyle{spmpsci.bst}
% \bibliography{./bibfile}

\begin{thebibliography}{10}
\providecommand{\url}[1]{{#1}}
\providecommand{\urlprefix}{URL }
\expandafter\ifx\csname urlstyle\endcsname\relax
  \providecommand{\doi}[1]{DOI~\discretionary{}{}{}#1}\else
  \providecommand{\doi}{DOI~\discretionary{}{}{}\begingroup \urlstyle{rm}\Url}\fi

\bibitem{Bergshoeff:2002qk}
Bergshoeff, E., Cucu, S., De~Wit, T., Gheerardyn, J., Halbersma, R., Vandoren, S., Van~Proeyen, A.: {Superconformal N=2, D = 5 matter with and without actions}.
\newblock JHEP \textbf{10}, 045 (2002).
\newblock \doi{10.1088/1126-6708/2002/10/045}

\bibitem{Bergshoeff:2004kh}
Bergshoeff, E., Cucu, S., de~Wit, T., Gheerardyn, J., Vandoren, S., Van~Proeyen, A.: {N = 2 supergravity in five-dimensions revisited}.
\newblock Class. Quant. Grav. \textbf{21}, 3015--3042 (2004).
\newblock \doi{10.1088/0264-9381/23/23/C01}

\bibitem{Bergshoeff:2001hc}
Bergshoeff, E., de~Wit, T., Halbersma, R., Cucu, S., Derix, M., Van~Proeyen, A.: {Weyl multiplets of N=2 conformal supergravity in five-dimensions}.
\newblock JHEP \textbf{06}, 051 (2001).
\newblock \doi{10.1088/1126-6708/2001/06/051}

\bibitem{Buchbinder:1998qv}
Buchbinder, I.L., Kuzenko, S.M.: {Ideas and methods of supersymmetry and supergravity: Or a walk through superspace} (1998)

\bibitem{Butter:2009cp}
Butter, D.: {N=1 Conformal Superspace in Four Dimensions}.
\newblock Annals Phys. \textbf{325}, 1026--1080 (2010).
\newblock \doi{10.1016/j.aop.2009.09.010}

\bibitem{Butter:2016mtk}
Butter, D., Ciceri, F., de~Wit, B., Sahoo, B.: {Construction of all N=4 conformal supergravities}.
\newblock Phys. Rev. Lett. \textbf{118}(8), 081,602 (2017).
\newblock \doi{10.1103/PhysRevLett.118.081602}

\bibitem{Butter:2010jm}
Butter, D., Kuzenko, S.M.: {New higher-derivative couplings in 4D N = 2 supergravity}.
\newblock JHEP \textbf{03}, 047 (2011).
\newblock \doi{10.1007/JHEP03(2011)047}

\bibitem{Butter:2013rba}
Butter, D., Kuzenko, S.M., Novak, J., Tartaglino-Mazzucchelli, G.: {Conformal supergravity in three dimensions: Off-shell actions}.
\newblock JHEP \textbf{10}, 073 (2013).
\newblock \doi{10.1007/JHEP10(2013)073}

\bibitem{Butter:2014xxa}
Butter, D., Kuzenko, S.M., Novak, J., Tartaglino-Mazzucchelli, G.: {Conformal supergravity in five dimensions: New approach and applications}.
\newblock JHEP \textbf{02}, 111 (2015).
\newblock \doi{10.1007/JHEP02(2015)111}

\bibitem{Butter:2018wss}
Butter, D., Novak, J., Ozkan, M., Pang, Y., Tartaglino-Mazzucchelli, G.: {Curvature squared invariants in six-dimensional ${\cal N} = (1,0)$ supergravity}.
\newblock JHEP \textbf{04}, 013 (2019).
\newblock \doi{10.1007/JHEP04(2019)013}

\bibitem{Butter:2017jqu}
Butter, D., Novak, J., Tartaglino-Mazzucchelli, G.: {The component structure of conformal supergravity invariants in six dimensions}.
\newblock JHEP \textbf{05}, 133 (2017).
\newblock \doi{10.1007/JHEP05(2017)133}

\bibitem{Butter:2013lta}
Butter, D., de~Wit, B., Kuzenko, S.M., Lodato, I.: {New higher-derivative invariants in N=2 supergravity and the Gauss-Bonnet term}.
\newblock JHEP \textbf{12}, 062 (2013).
\newblock \doi{10.1007/JHEP12(2013)062}

\bibitem{Casarin:2024qdn}
Casarin, L., Kennedy, C., Tartaglino-Mazzucchelli, G.: {Conformal anomalies for (maximal) 6d conformal supergravity}  (2024)

\bibitem{Freedman:2012zz}
Freedman, D.Z., Van~Proeyen, A.: {Supergravity}.
\newblock Cambridge Univ. Press, Cambridge, UK (2012).
\newblock \doi{10.1017/CBO9781139026833}

\bibitem{Fujita:2001kv}
Fujita, T., Ohashi, K.: {Superconformal tensor calculus in five-dimensions}.
\newblock Prog. Theor. Phys. \textbf{106}, 221--247 (2001).
\newblock \doi{10.1143/PTP.106.221}

\bibitem{Gates:1983nr}
Gates, S.J., Grisaru, M.T., Rocek, M., Siegel, W.: {Superspace Or One Thousand and One Lessons in Supersymmetry}, \emph{Frontiers in Physics}, vol.~58 (1983)

\bibitem{PRL}
Gold, G., Hutomo, J., Khandelwal, S., Ozkan, M., Pang, Y., Tartaglino-Mazzucchelli, G.: All gauged curvature-squared supergravities in five dimensions.
\newblock Phys. Rev. Lett. \textbf{131}, 251,603 (2023).
\newblock \doi{10.1103/PhysRevLett.131.251603}.
\newblock \urlprefix\url{https://link.aps.org/doi/10.1103/PhysRevLett.131.251603}

\bibitem{Gold:2023ykx}
Gold, G., Hutomo, J., Khandelwal, S., Tartaglino-Mazzucchelli, G.: {Components of curvature-squared invariants of minimal supergravity in five dimensions}  (2023)

\bibitem{Gold:2023dfe}
Gold, G., Hutomo, J., Khandelwal, S., Tartaglino-Mazzucchelli, G.: {Curvature-squared invariants of minimal five-dimensional supergravity from superspace}.
\newblock Phys. Rev. D \textbf{107}(10), 106,013 (2023).
\newblock \doi{10.1103/PhysRevD.107.106013}

\bibitem{GoldGithub}
Gold, G., Khandelwal, S., Tartaglino-Mazzucchelli, G.: Sugra component reduction.
\newblock \url{https://github.com/gregory-gold/sugra-component-reduction} (2024)

\bibitem{Hanaki:2006pj}
Hanaki, K., Ohashi, K., Tachikawa, Y.: {Supersymmetric Completion of an R**2 term in Five-dimensional Supergravity}.
\newblock Prog. Theor. Phys. \textbf{117}, 533 (2007).
\newblock \doi{10.1143/PTP.117.533}

\bibitem{Kugo:2000hn}
Kugo, T., Ohashi, K.: {Supergravity tensor calculus in 5-D from 6-D}.
\newblock Prog. Theor. Phys. \textbf{104}, 835--865 (2000).
\newblock \doi{10.1143/PTP.104.835}

\bibitem{Kugo:2000af}
Kugo, T., Ohashi, K.: {Off-shell D = 5 supergravity coupled to matter Yang-Mills system}.
\newblock Prog. Theor. Phys. \textbf{105}, 323--353 (2001).
\newblock \doi{10.1143/PTP.105.323}

\bibitem{Kugo:2002vc}
Kugo, T., Ohashi, K.: {Gauge and nongauge tensor multiplets in 5-D conformal supergravity}.
\newblock Prog. Theor. Phys. \textbf{108}, 1143--1164 (2003).
\newblock \doi{10.1143/PTP.108.1143}

\bibitem{Kuzenko:2015jda}
Kuzenko, S.M., Novak, J., Tartaglino-Mazzucchelli, G.: {Higher derivative couplings and massive supergravity in three dimensions}.
\newblock JHEP \textbf{09}, 081 (2015).
\newblock \doi{10.1007/JHEP09(2015)081}

\bibitem{Kuzenko:2022ajd}
Kuzenko, S.M., Raptakis, E.S.N., Tartaglino-Mazzucchelli, G.: {Covariant Superspace Approaches to $\mathcal{N}=\textrm{2}$ Supergravity} (2023)

\bibitem{Kuzenko:2022skv}
Kuzenko, S.M., Raptakis, E.S.N., Tartaglino-Mazzucchelli, G.: {Superspace Approaches to $\mathcal{N}=\textrm{1}$ Supergravity} (2023)

\bibitem{MacCallum:2018}
MacCallum, M.: Computer algebra in gravity research.
\newblock Living Reviews in Relativity \textbf{21} (2018).
\newblock \doi{10.1007/s41114-018-0015-6}

\bibitem{Novak:2017wqc}
Novak, J., Ozkan, M., Pang, Y., Tartaglino-Mazzucchelli, G.: {Gauss-Bonnet supergravity in six dimensions}.
\newblock Phys. Rev. Lett. \textbf{119}(11), 111,602 (2017).
\newblock \doi{10.1103/PhysRevLett.119.111602}

\bibitem{Ozkan:2024euj}
Ozkan, M., Pang, Y., Sezgin, E.: {Higher Derivative Supergravities in Diverse Dimensions}  (2024)

\bibitem{Peeters:2006kp}
Peeters, K.: {A Field-theory motivated approach to symbolic computer algebra}.
\newblock Comput. Phys. Commun. \textbf{176}, 550--558 (2007).
\newblock \doi{10.1016/j.cpc.2007.01.003}

\bibitem{peeters2007introducing}
Peeters, K.: Introducing cadabra: a symbolic computer algebra system for field theory problems (2007)

\bibitem{CadabraGithub}
Peeters, K.: Cadabra.
\newblock \url{https://github.com/kpeeters/cadabra2} (2015)

\bibitem{Peeters:2018dyg}
Peeters, K.: {Cadabra2: computer algebra for field theory revisited}.
\newblock J. Open Source Softw. \textbf{3}(32), 1118 (2018).
\newblock \doi{10.21105/joss.01118}

\bibitem{Wess:1992cp}
Wess, J., Bagger, J.: {Supersymmetry and supergravity}.
\newblock Princeton University Press, Princeton, NJ, USA (1992)

\end{thebibliography}

\end{footnotesize}

\end{document}